%% file: isca06-cqla.tex
\documentclass[times,twocolumn,10pt]{article}
\usepackage{latex-ieee}

\input{macros}
\usepackage[]{epsfig}
\usepackage[]{multirow}
\usepackage[]{setspace}   
\usepackage[]{graphicx}
\usepackage[]{subfigure}  
\usepackage[]{dcolumn}    
\usepackage[]{bm}         
\usepackage[]{tightenum}  
\usepackage[]{amsmath}    
\usepackage[]{pslatex}    

\usepackage[]{setspace}   

\begin{document}
\title{\begin{footnotesize}To appear in the International Symposium on Computer Architecture (ISCA-33)\end{footnotesize}\\
Quantum Memory Hierarchies: Efficient Designs to Match Available
Parallelism in Quantum Computing}

\author{Darshan D. Thaker$^\dag$ \hspace{0.1cm} Tzvetan S. Metodi$^\dag$ \hspace{0.1cm} 
Andrew W. Cross$^\ddag$ \hspace{0.1cm} Isaac L. Chuang$^\ddag$  \hspace{0.1cm} Frederic T. Chong$^\star$\\ \\
$^\dag${\em University of California at Davis, One Shields Avenue, Davis, CA 95616, USA}\\
$^\ddag${\em Massachusetts Institute of Technology, 77 Massachusetts Avenue, Cambridge, MA 02139} \\
$^\star${\em University of California at Santa Barbara, Santa Barbara, CA 93106, USA}\\
}

\maketitle
\thispagestyle{empty}


\begin{abstract}
The assumption of maximum parallelism support for the successful
realization of scalable quantum computers has led to homogeneous,
``sea-of-qubits'' architectures. The resulting architectures
overcome the primary challenges of reliability and scalability at
the cost of physically unacceptable system area. We find that by
exploiting the natural serialization at both the application 
and the physical microarchitecture level of a quantum computer, we
can reduce the area requirement while improving 
performance. In particular we present a scalable quantum
architecture design that employs specialization of the system into
memory and computational regions, each individually optimized to
match hardware support to the available parallelism. Through careful
application and system analysis, we find that our new architecture
can yield up to a factor of thirteen savings in area due to
specialization. In addition, by providing a memory hierarchy design
for quantum computers, we can increase time performance by a factor
of eight. This result brings us closer to the realization of a quantum
processor that can solve meaningful problems.
\end{abstract}

\vspace{-0.5cm}
\section{Introduction} \label{sec:intro}

\vspace{-0.5cm}

Conventional architectural design adheres to the concept of {\bf
balance}. For example, the register file depth is matched to the number of
functional units, the memory bandwidth to the cache miss rate, or
the interconnect bandwidth matched to the compute power of each
element of a multiprocessor.  We apply this concept to the design of
a quantum computer and introduce the {\it Compressed Quantum Logic
Array} (CQLA), an architecture that balances components and
resources in terms of exploitable parallelism. The primary goal of
our design is to address the problem of large area, approximately
$1$~m$^2$ on a side, of our previous design \cite{Metodi05}.

Specifically, we discover that the prevailing approach to designing
a quantum computer, that of supporting maximal parallelism, is area
inefficient. We also find that exploitable parallelism is
inherently limited by both resource constraints and application
structure. This lack of parallelism gives us the freedom to
increase density by specializing components as blocks of
memory and blocks of computation.

We introduce the idea of periodically reducing our investment in
reliability and thereby increasing speed. By encoding the compute
regions differently than memory we provide very fast compute
regions, while allowing the memory to be slower and more reliable.
To ensure that the faster compute region does not suffer from too
many stalls, we employ a quantum memory hierarchy wherein the cache
utilizes the same encoding mechanism as the compute region. When
making this effort to improve speed, it is critical that overall
system fidelity is maintained. We show how this can be accomplished.

Due to the quantum no-cloning theorem \cite{Zurek}, it is necessary
for all quantum data to physically move from source to destination.
We cannot create a copy of the data and send the copy. Our
architecture focuses on implementation with an array of trapped
atomic ions, one of the most mature and scalable technologies that
provides a wealth of experimental data. In ion-traps, the physical
representation of data are ions that are in constant motion, on a
two dimensional grid, throughout the computation. Since this
physical movement is slow, yet unavoidable, it limits available
parallelism at the microarchitecture level.

At the application level, we find that only a limited amount of
parallelism can be extracted from key quantum algorithms. This means
that we may only need a few compute blocks for all the qubits in
memory. This is in contrast to the popular ``sea of qubits'' model
which allows computation at every qubit.  Our results show up to a
13X increase in density, particulary important in addressing our
primary goal, and a speedup of about $8$. The large area improvement
brings the engineering of a quantum architecture closer to the
capabilities of current implementation technologies.

The choice of quantum error correction codes (ECC) influences our
results and the architecture. In our specialized architecture
analysis, we use the previously considered Steane $[[7,1,3]]$ code
\cite{Steane97a} and utilize a newly optimized Bacon-Shor
$[[9,1,3]]$ code \cite{Bacon05,Poulin05}. The \bacon code, though
larger than the \steane code since it uses more physical qubits to
encode a single logical qubit, requires far fewer resources for
error-correction \cite{Aliferis05b}, thus reducing the overall area
and increasing the speed.

Furthermore, we find that
communication is generally dominated by
computation for error correction. This computation allows us to
absorb the cost of moving data between different regions of the
architecture. Error correction is so substantial, in fact,
that quantum computers do not suffer from the {\it memory wall} faced by
conventional computers. Thus our dense structure with a
communication infrastructure based on our prior work \cite{Metodi05}
can accommodate applications with highly-demanding communication
patterns.

In summary, the {\bf contributions} of this work are: 1) Our specialized
architecture, the CQLA, successfully tackles the issue of size,
which has been the biggest drawback facing large-scale realizable
quantum computers. 2) We show that current parallelism in
quantum algorithms is inherently limited and consideration of
physical resources and data movement restrict it even further. 3) We
present and analyze the abstractions of memory, cache and
computation units for a quantum computer; based on the insight that
we can reduce reliability for the compute units and cache without
sacrificing overall computation fidelity. This approach helps us
significantly increase the performance of the system.

The paper is organized as follows. Section \ref{sec:background}
provides a background of
the homogeneous QLA architecture and the low-level microarchitecture
assumptions of our system. Section \ref{sec:abstractions} motivates
the specialized CQLA architecture and introduces the architectural
abstractions. Thereafter we discuss how the Steane and
Bacon-Shor error correction codes affect the design of  the
CQLA. Results and analysis of our abstractions are the focus of section
\ref{sec:cqla_optim} following which we provide details of
computation versus
communication requirements of the most widely accepted quantum
applications. We end with future directions in Section \ref{sec:issues}
and our conclusions in Section \ref{sec:conclude}.

\section{Background} \label{sec:background}

\vspace{-0.5cm}
Our architectural model is built upon our previous work on the
Quantum Logic Array (QLA) architecture \cite{Metodi05}. The QLA
architecture is a hierarchical array-based design that overcomes the
primary challenges of scalability for large-scale quantum
architectures. It is a homogeneous, tiled architecture with three
main components: logical qubits implemented as self-contained
computational tiles structured for quantum error error correction;
trapped atomic ions as the underlying technology; finally,
teleportation-based communication channels utilizing the concept of
quantum repeaters to overcome the long-distance communication
constraints.

\subsection{The Logical Qubit}
\vspace{-0.5cm}

The basic structure of the QLA, our prior work,
implements a fault-tolerant quantum
bit, or a {\em logical qubit} as a self-contained tile whose
underlying construction is intended for quantum error correction, by
far the most dominant and basic operation in a quantum machine
\cite{Oskin02}. Quantum error correction is expensive because
arbitrary reliability is achieved by recursively encoding physical
qubits at cost of exponential 
overhead. Recursive error correction works by encoding $N$ physical
ion-qubits into a known highly-correlated state that can be used to
represent a single logical data qubit. This data qubit is now at
level $1$ recursion and may have the property of being in a
superposition of ``0'' and ``1'' much like a single physical qubit.
Encoding once more we can create a logical qubit at level $2$
recursion with $N^2$ physical ion-qubits. With each level, $L$, of
encoding the probability of failure of the system scales as
$p_{0}^{2^L}$, where $p_0$ is the failure rate of the individual
physical components given a fault-tolerant arrangement and sequence
of operations for the lower level components. The ability to apply
logical operations on a logical qubit without the need to decode and
subsequently re-encode the data is key to the existence of
fault-tolerant quantum microarchitecture design, where arbitrary
reliability can be efficiently reached through recursive encoding.

\begin{figure*}
\centering \subfigure[]{
    \label{fig:ion_trap:a}
    \includegraphics[width=2.4in]{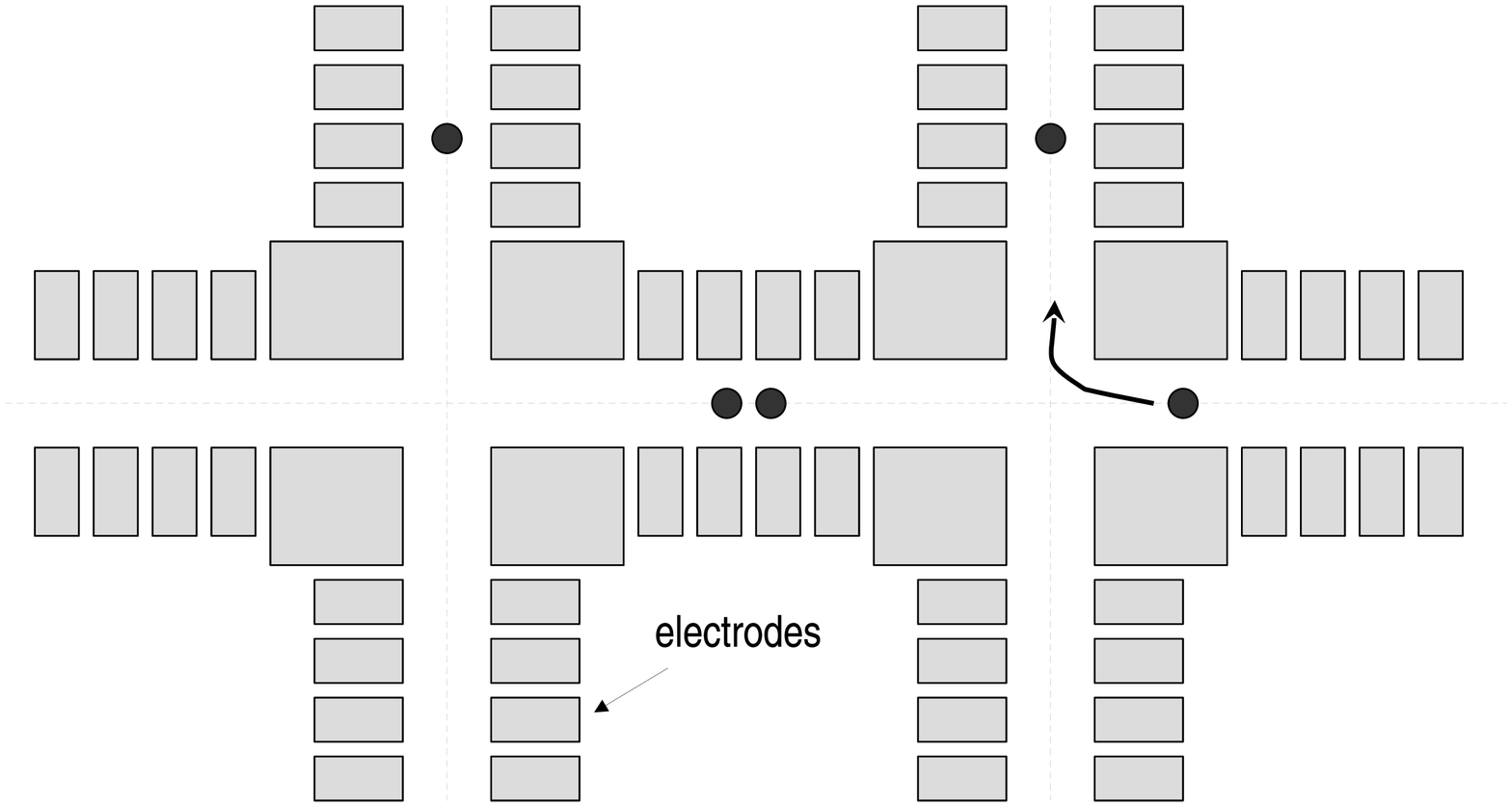}}
\hspace{1cm} \centering \subfigure[]{
    \label{fig:ion_trap:b}
    \includegraphics[width=2.2in]{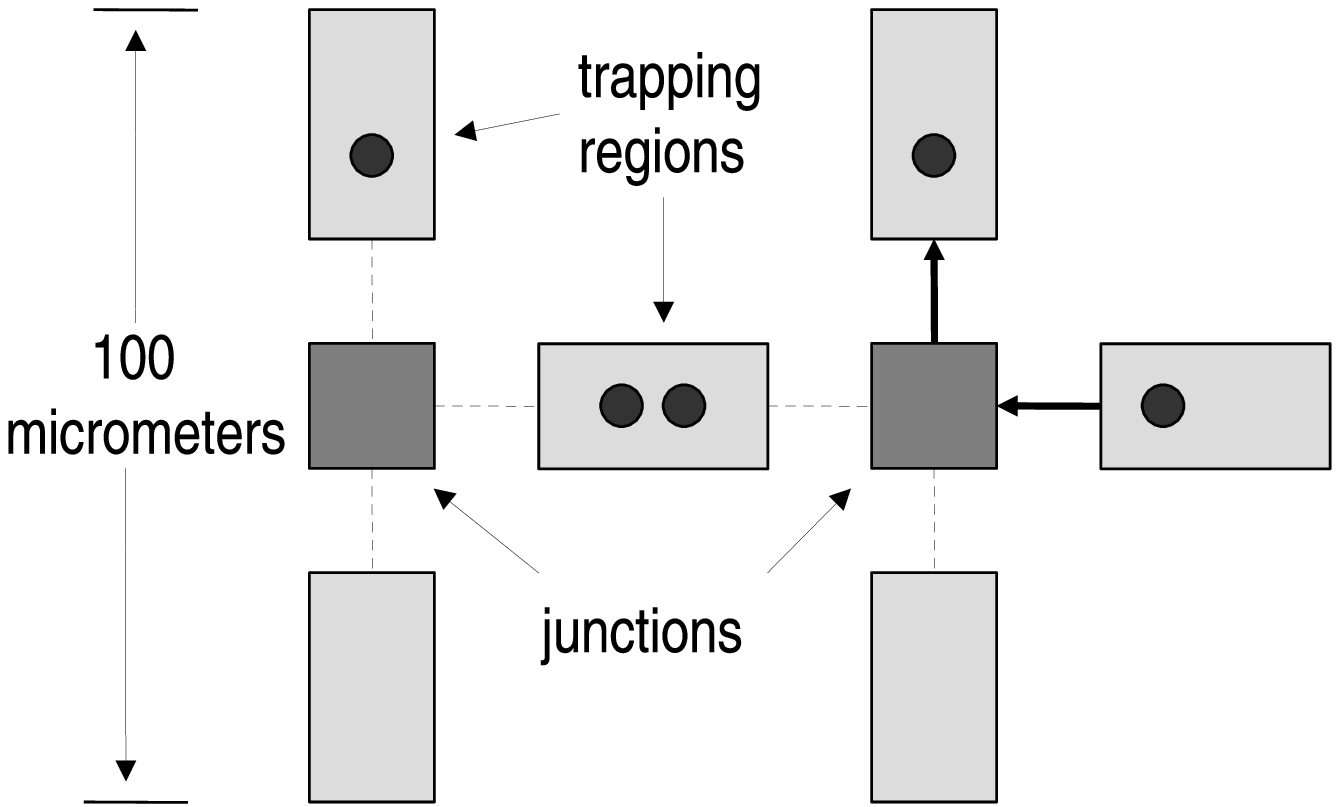}}
\caption{\textsf{(a) A simple schematic of the basic elements of a
planar ion-trap for quantum computing. Ions are trapped in any of
the trapping regions shown and ballistically shuttled from one
trapping region to another. When two ions are together a two-qubit
gate can be performed. (b) Our abstraction of the ion-trap layout.
Each trapping region can hold up to two ions for two-qubit gates.
The trapping regions are interconnected with the crossing junctions
which are treated as a shared resource.}} \label{fig:ion_trap}
\end{figure*}

The logical qubits in the QLA are arranged in a regular array
fashion, connected with a tightly integrated repeater-based
\cite{Dur98a} interconnect. This makes the high-level design of the
QLA very similar to classical tile based architectures. The key
difference is that the communication paths must account for data
errors in addition to latency. Integrated repeaters known as {\em
teleportation islands} redirect qubit traffic in the $4$ cardinal
directions by teleporting data from one island to the next. This
interconnect design is one of the key innovative features of QLA
architecture, as it allows us to completely overlap communication
and computation, thus eliminating communication latency at the
application level of the program.

Anticipating technology improvements in the near future we found
that for performing large, relevant instances of Shor's factoring
algorithm, sufficient reliability is achieved at level $2$ encoding
per logical qubit using the Steane \steane error correction
code \cite{Steane96}. In the QLA, computation could occur at any
logical qubit and each logical gate is followed by an error correction
procedure.  To preserve homogeneity and maximum flexibility for large-scale
applications each logical qubit was accompanied by all necessary
error correction auxiliary qubit resources such that computational
speed was maximized.  This amounted to a $(1:2)$ ratio between logical
data qubits and ancillary qubits.

\subsection{Low-Level Physical Architecture Model}
\label{sec:traps}
\vspace{-0.5cm}

At the lowest level our architecture design is based on the ion-trap
technology for quantum computation. Initially proposed by Cirac and
Zoller in 1995 \cite{Cirac95}, the technology uses a number of
atomic ions that interact with lasers to quantum compute. Quantum
data is stored in the internal electronic and nuclear states of the ions, 
while
the traps themselves are segmented metal traps (or electrodes) that
allow individual ion addressing. Two ions in neighboring traps can
couple to each other forming a linear chain of ions whose
vibrational modes provide qubit-qubit interaction used for
multi-qubit quantum gates \cite{Sorensen00,Leibfried03}. Together
with single bit rotations this yields a universal set of quantum
logic gates. All quantum logic is implemented by applying lasers on
the target ions, including measurement of the quantum state
\cite{Hahn50,Wineland98,Kielpinski02,Williams03a}.  Sympathetic
cooling ions absorb vibrations from data ions, which are then
dampened through laser manipulation \cite{Kielpinski99,Blinov02}.
Recent experiments \cite{Barrett04,Riebe04,Chiaverini05} have
demonstrated all the necessary components needed to build a
large-scale ion-trap quantum information processor.  Finally,
multiple ions in different traps can be controlled by focusing
lasers through MEMS mirror arrays \cite{Kim05}.

Figure \ref{fig:ion_trap} shows a schematic of the physical
structure of an ion trap computer element. In Figure
\ref{fig:ion_trap:a} we see a single ion trapped in the middle
trapping region. Trapping regions are the locations where ions can
be prepared for the execution of a logical gate, which is implemented
by an external laser source pulsed on the ions in the trap. In the figure
we see an ion moving from the far right trapping region to the
top-right for the execution of a two-bit logical operation.

\begin{table}
    \begin{center}
    \begin{footnotesize}
    \begin{tabular}{|l|l|l|}
        \hline
        Operation & Time $\mu$s now(future)  & Failure Rate now(future)\\
        \hline 
                     &                   &                                         \\
        Single Gate  &  $1~(1)$      &   $10^{-4}~(10^{-8})$       \\
        Double Gate  &  $10~(10)$    &   $0.03~(10^{-7})$       \\
        Measure      &  $200~(10)$   &   $0.01~(10^{-8})$       \\
        Movement     &  $20~(10)$   &   $0.005~(5\times 10^{-8})/\mu$m \\
        Split        &  $200~(0.1)$     &                               \\
        Cooling      &  $200~(0.1)$      &                                          \\
        \hline
        Memory time  &  $10$~to~$100$ sec &                                 \\
        Trap Size    &  $\sim 200~(1-5)~\mu$m &\\
        \hline
    \end{tabular}
    \end{footnotesize}
    \end{center}
    \caption{\textsf{Column $1$ gives estimates for execution times for basic physical
    operations used in the QLA model.  Currently achieved
    component failure rates are based on experimental measurements
    at NIST with $^9Be^+$ ions, and using $^{24}Mg^+$ ions for sympathetic
    cooling \cite{Wineland98,Leibfried03}. All parameters are followed by their projected
    parameters in parenthesis, extrapolated following recent literature
    \cite{ARDA,Steane04b,Ozeri05}, and discussions with the NIST researchers; these
    estimates are used in modeling the performance of our architecture.}}
    \vspace{-0.3in} \label{table:params}
\end{table}

Figure \ref{fig:ion_trap:b} demonstrates our abstraction of the
physical ion-trap layout. The layout can be represented as a
collection of trapping regions connected together through shared
junctions. A fundamental time-step, or a clock cycle, in an ion-trap
computer will be defined as any physical, unencoded logic operation 
(one-bit or two-bit), a basic move operation from one trapping region to
another, and measurement. Table \ref{table:params} summarizes
current experimental parameters and corresponding optimistic
parameters for ion-traps. In our subsequent analysis we will assume
that each {\em clock cycle} for a fundamental time-step has a
duration of $10~\mu$s, failure rates are $10^{-8}$ for single-qubit
operations and measurement, $10^{-7}$ for \cnot gates
\cite{Ozeri05}, and order of $10^{-6}$ per fundamental move operation. The
movement failure rate is expected to improve from what it is now as
trap sizes shrink and electrode surface integrity continues to
improve. We will assume trap sizes of $5\mu$m each
\cite{Wineland05}, and on the order of $10$ electrodes per trapping region
\cite{Hensinger05}, which gives us a trapping region dimension
(including the junction) of $50\mu$m. The parameters chosen for our
study are optimistic compared to \cite{Oskin05a} and \cite{Meter05}.
Both of those papers, assume more pessimistic near term parameters
which are useful for building a 100 bit prototype, but probably not a
scalable quantum computer that can factor 1024-bit numbers using
Shor's algorithm. Based on the quantum computing ARDA roadmap
\cite{ARDA}, we feel justified in using aggressive parameters when
looking 10-15 years into the future.


\section{Architectural Abstractions} \label{sec:abstractions}

\vspace{-0.5cm}

\begin{figure}
\centering{
\includegraphics[width=3.0 in]{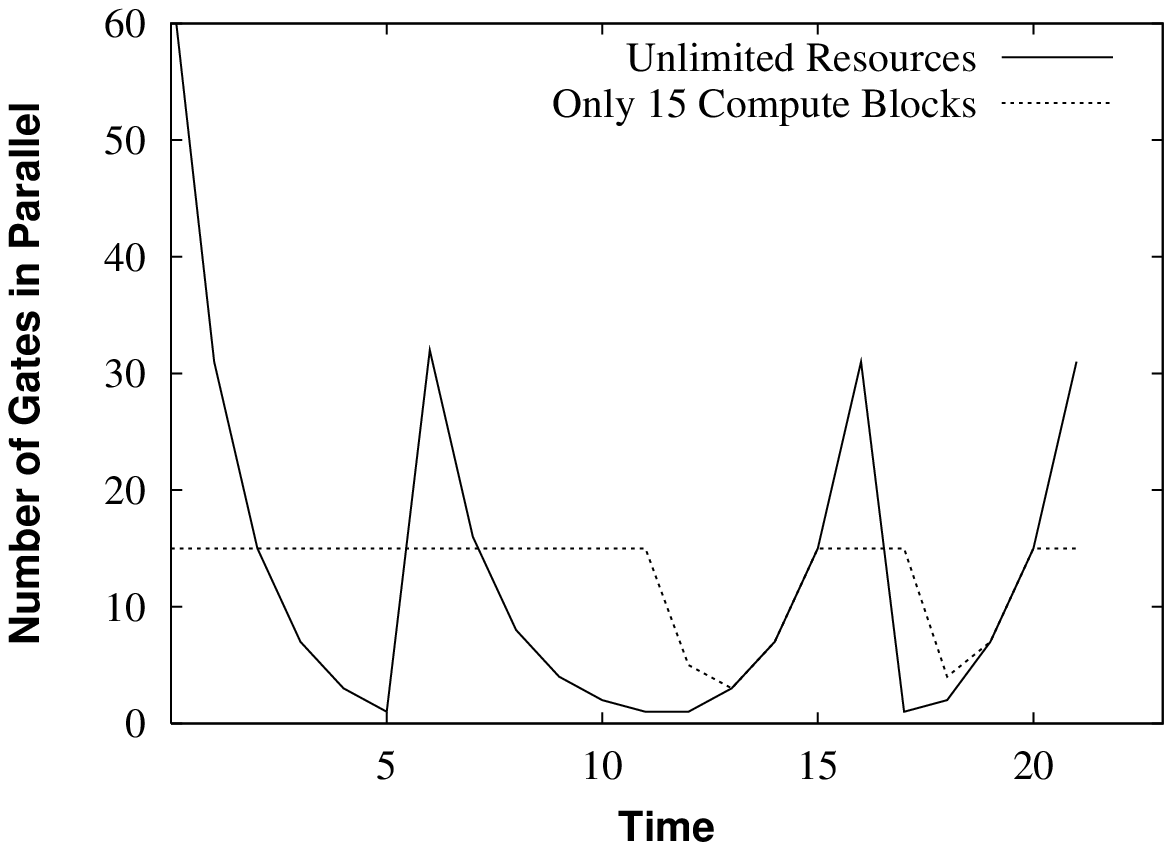}}
\caption{\textsf{For a $64$-qubit adder, the amount of parallelism
that can be extracted when resources are unlimited, and when the
number of gates per cycle are limited. This figure shows that if 15
gates, or an unlimited number of gates could be performed in each
cycle, the total runtime would remain the same. compute blocks
increases.}}\label{fig:algo_parallel}
\end{figure}

This section motivates the need for a compact architecture for quantum
processors and describes our design the CQLA (Compressed Quantum Logic Array).
We discuss how separation into memory and compute regions benefits
the CQLA and then present our quantum memory hierarchy.

\subsection{Motivation}
\vspace{-0.5cm}
Conventional quantum processor designs are based on the 
{\em sea-of-qubits} design and allow computation to take place
anywhere in the processor. This design philosophy follows the idea of
maximum parallelism and is employed in our previous work
\cite{Metodi05}. The area consumption of such a design however,
is untenably large, about $1$ m$^2$ to factor a $1024$-bit number.

When we consider the amount of available parallelism in quantum
applications, we discover that much is to be gained by limiting
computation to a specifically designated location. The remaining
area can be optimized for storage of quantum data. A good example
for the benefit of specialization in quantum applications is the
Draper carry-lookahead quantum adder \cite{Draper04}, which forms
a basic basic component of Shor's quantum factoring algorithm
\cite{Shor94}.  Figure \ref{fig:algo_parallel} shows that providing
unlimited computational resources for a $64$-bit 
adder does not offer a performance benefit over limiting 
the computation to $15$ locations.  As illustrated in Section
\ref{sec:background}, the number of ancillary resources for each
data location where computation is allowed is twice as large. In this
example, by providing only $15$ compute locations instead of $64$, we
can reduce the area consumed by the adder by approximately {\em half}
and yet have no change in performance.

\begin{figure*}
\centering \subfigure[]	{
    \label{fig:design_a}
    \includegraphics[width=2.7in]{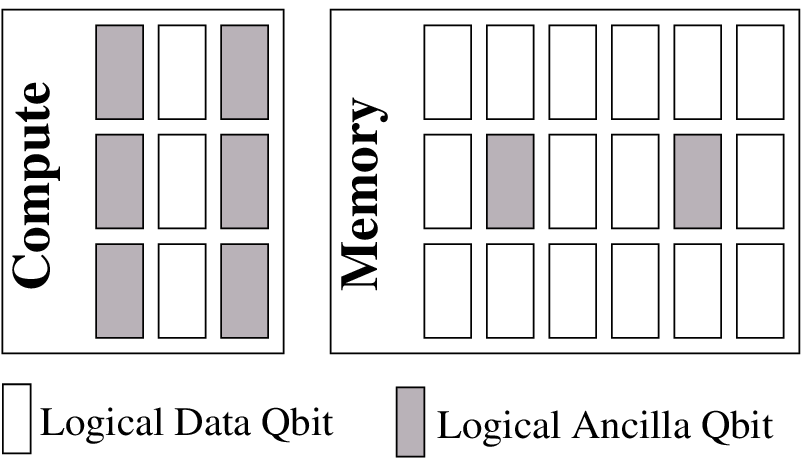}}
\hspace{1cm}\centering \subfigure[]{
    \label{fig:design_b}
    \includegraphics[width=2.7in]{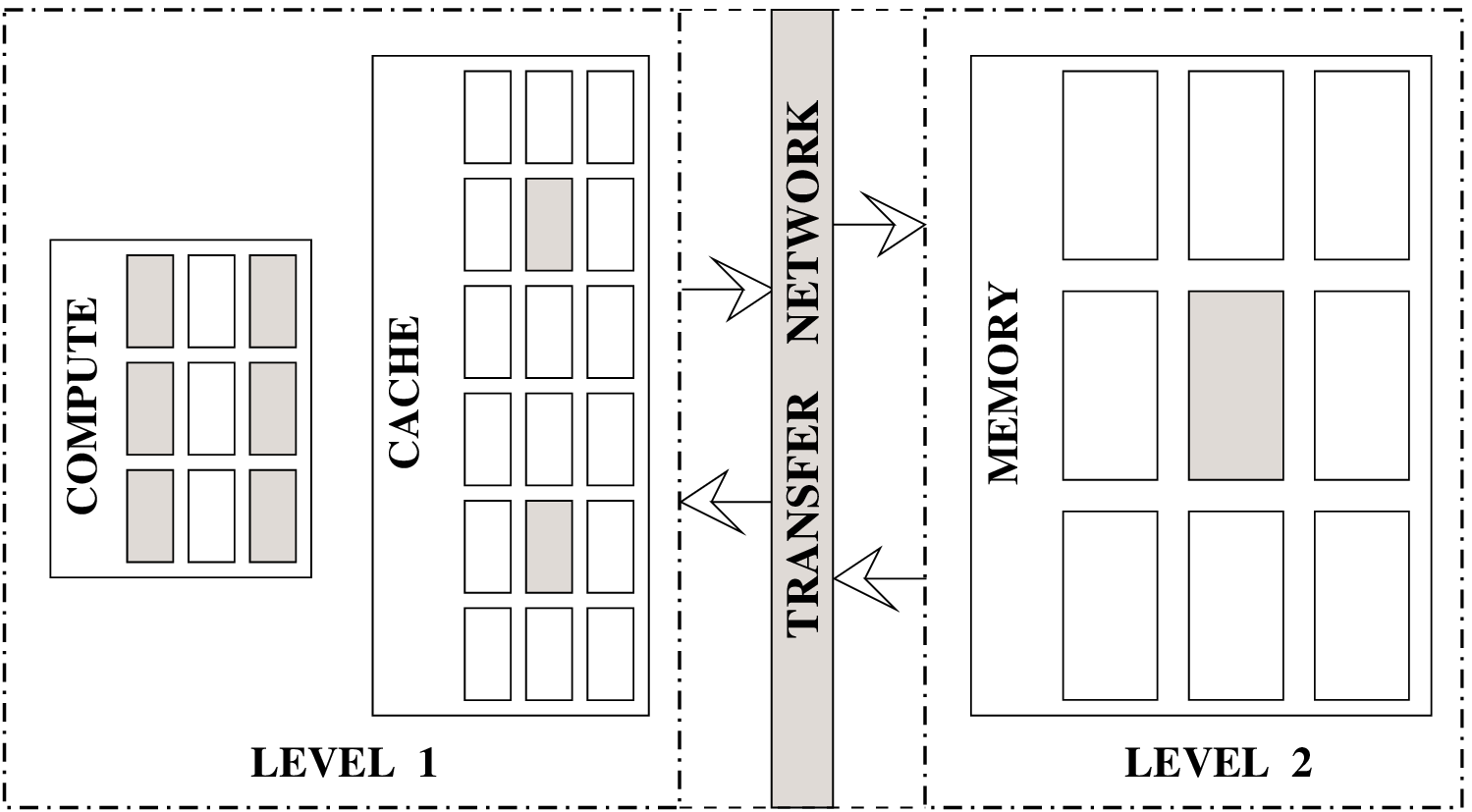}}
    \caption{\textsf{{\bf (a)}
    Memory is denser since it has fewer ancilla qubits. 
    The figure shows $3$ data qubits
    in the compute block which take the same area as $8$ data qubits in
    memory. In the CQLA each compute block holds nine
    $9$ data qubits and $18$ ancilla. Both compute and memory are at level 2
    encoding. {\bf (b)} Memory is at level $2$ encoding, while the compute
    and cache are at level $1$ encoding. The complete CQLA consists of memory
    at level $2$, compute regions at level $2$ and also a cache and compute
    region at level $1$.}}
    \label{fig:design}
\end{figure*}

\subsection{Specialized Components} \label{sec:special}
\vspace{-0.5cm}
The facts that qubits in an ion-trap quantum processor have {\em large
lifetimes} when idle, allows us to improve logical
qubit density in the memory. 
Qubits in memory can wait for a longer time period between two
consecutive error corrections. We use this to significantly reduce
the error correction ancillary resources in memory, 
thereby reducing its density.
The majority of computation, on the other hand,
is an interaction between two distinct logical qubits. 
To maintain adequate system fidelity, every gate must be
followed by an error correction procedure. Consequently, a quantum
processor spends most of its time performing error correction and
the compute regions are designed to allow fast error
correction by providing a greater number of ancilla in the logical
qubits.  Figure 
\ref{fig:design_a} shows a specialization into compute and memory
regions. The ratio of (data:ancilla) can be seen to be $(8:1)$ for
memory and $(1:2)$ for the compute region.

While specialization helps address our primary goal of reducing
size, it can possibly also reduce performance. In Section \ref{sec:cqla_optim}
we show how judiciously choosing the size of the compute region
helps maintain adequate performance while simultaneously reducing size .

\subsection{Quantum Memory Hierarchy}\label{sec:mem_h}
\vspace{-0.5cm}
Another important architectural design choice is the effect of the
error correction code chosen in both the memory and the compute
regions. Error correction is the most dominant procedure and the
resources used increase exponentially with each level of
concatenation. In addition to resources, the time to error correct
increases exponentially with each level of concatenation. The
benefits of concatenated error correction are that the reliability
of each operation increases double exponentially, thus allowing far
greater number of total operations to be performed. 
For any application,  all logical qubits are not being acted
upon by gates for the entire duration of the algorithm. In fact,
just like classical computers, data locality is a common phenomenon. 
This implies that a logical qubit could start at level 2 encoding,
be encoded at level 1 during the peak in its activity and return
to level 2 when idle. 

We now introduce a quantum memory hierarchy, 
in addition to the specialized design.  Memory at level $2$,
which is optimized for
area and reliability will be inherently slower than a computational
structure, at level $1$, optimized for gate execution. 
This necessitates the need for
a cache that can alleviate the need for constant communication.

Figure \ref{fig:design_b} outlines this approach.
the separation between memory and compute regions. 
The cache and the compute
regions here are similar to Figure \ref{fig:design_a} in every way
save that they are at a lower level of encoding.  In the memory hierarchy,
memory and cache have a similar design, only memory is at a higher level
of encoding, and hence is slower and much more reliable.  The critical feature
here is the {\bf transfer network} which is more complicated and hence
slower than the teleporation channels described above. The transfer network
comes into play only when we change the encoding of a logical qubit. For
all other communication (within compute blocks,  between cache and compute
blocks and within memory) teleportation is still the chosen mechanism.
Section \ref{sec:ECC} describes how the transfer process is performed
in a fault-tolerant manner.

\section{Error Correction and Code Transfer}\label{sec:ECC}
\vspace{-0.5cm}

In this section we describe the cost of the
error correction circuits and code-transfer networks we use when a
specific physical layout is considered. Section \ref{sec:traps}
describes in detail our technology parameters, which we find to be
necessary for such a large-scale architecture. These parameters
allow the large scalability to be achieved because the physical
component failure rates are below the threshold value needed for
efficient error correction \cite{Aharonov97a}.

\subsection{Error Correction Codes}
\vspace{-0.5cm}
Some of the best error correction codes (ECC) are ones that use
very few physical qubits, and allow ``easy'' fault-tolerant gate
implementations. A requirement of a fault-tolerant system is that
computation proceeds without decoding the encoded data. Thus logical
gates are implemented directly on encoded qubits, ensuring that
errors introduced during the gate can be corrected. Many code
choices for EC allow {\em transversal} logical gate implementation,
which means that the same physical gate acts on each lower-level
qubit. 

Each logical quantum gate is preceded and followed by an error
correction procedure. The EC procedure works by encoding ancillary
qubits in the logical ``0'' state of the data and interacting the
data and the ancilla. The interaction causes errors in the data to
propagate to the ancilla and to be detected when the ancilla is
measured. There are several very important logical gates that we
must consider during error correction. The bit-flip gate, $X$ flips
the value of the qubit by reversing the probabilities between its
``0'' component and its ``1'' component. The phase-flip gate, $Z$,
acts only on the qubit's ``1'' component by changing its sign. The
most important gate is the controlled-$X$ gate (denoted as the \cnot
gate) which flips the state of the target qubit whenever the state
of the control qubit is set. Errors on the data can be understood as
the product of phase-flips and a bit-flips. A syndrome is extracted
for each types of error. We only present the cost of error
correction networks and details relevant to building a large-scale
architecture. The interested reader can refer to the literature for
additional theoretical information \cite{Nielsen00a}.

\begin{figure}[htbp]
\centering{
\includegraphics[width=3in]{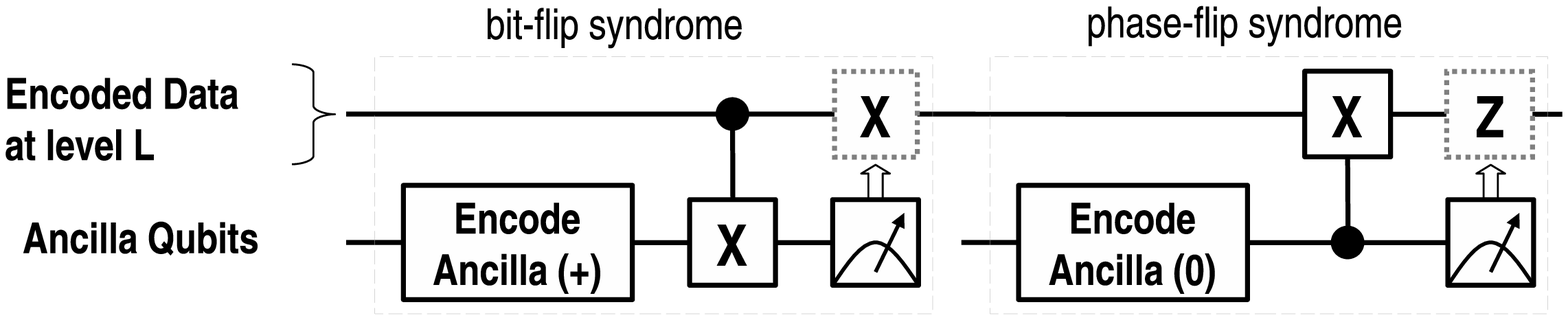}
\caption{\textsf{A high-level view of an error correction sequence.
Two syndromes for bit-flip and phase-flip errors are
extracted.}}\label{fig:ec_high_level}}
\end{figure}

Figure \ref{fig:ec_high_level} is a simple schematic of the general
error correction procedure, where time flows from left to right and
each line represents the evolution of an encoded logical qubit. An
error correction code is labeled by \ecc,
encoding $k$ logical qubits into $n$ qubits and 
correcting $(d-1)/2$ errors. If our target
reliability is such that we require $L$ levels of recursion, each
line in Figure \ref{fig:ec_high_level} represents $n^L$ level zero
qubits. For the bit-flip error syndrome the ancilla are encoded into
the logical $(0 + 1)$, and the transversal \cnot gate, which is
essentially $n$ level $(L-1)$ transversal \cnot gates of which the
ancillary qubits are targets. Each of the lower level \cnot gates is
followed by a lower level error correction unless the lower level is
zero. In our architecture analysis we provide information about two
error correcting codes: the Steane \steane code \cite{Steane96}, and
an improved version of the Shor \bacon code \cite{Shor95} denoted
as the Bacon-Shor code \cite{Bacon05,Poulin05,Aliferis05b}.

\noindent{\bf The Steane \steane Code} encodes $1$ qubit into $7$
qubits, and is the smallest error correction code allowing
transversal gate implementation for all gates involved in concatenated
error correction algorithms. The addition of the $T$ phase gate, which
is harder to implement, provides universal quantum logic using
the \steane error correcting code. For this reason
it was used as the underlying error correcting code in the analysis 
of the QLA architecture \cite{Metodi05}. 
It consists of $7$ data ions which encode our
logical level $1$ qubit with $14$ ancillary ions used for error correction,
seven of which are used in the error correction and the other verify
the ancilla.

Considering communication, the level $1$ error correction circuit in
will take $154$ cycles, where each cycle is in the order of
$10$ microseconds, and
can be as large as $0.003$ per error correction procedure at level 1.
A level 2 \steane qubit will be composed of $7$
level $1$ data qubits and $7$ level $1$ ancilla qubits - there is no need for
verification ancilla at $L=2$. The size of a level $2$ qubit will be
$3.4~mm^2$, and a fully serialized error correction will last
approximately $0.3$ seconds (this is two orders of magnitude more than
the time to error correct at level 1).

\noindent{\bf Bacon-Shor \bacon Code:} The \bacon code was
the first error correcting code to be discovered for arbitrary
errors \cite{Shor95}. Recent observations make this code faster and
spatially smaller than the \steane code \cite{Bacon05,Poulin05,Aliferis05b}. 
The compact
structure of the physical layout for the \bacon code significantly
improves communication requirements. At level 1 the error correction
time is only $0.001$ seconds and $0.1$ seconds at level
$2$. The level $2$ qubit size is approximately $2.4$~mm$^2$. Table
\ref{table:code_metrics} summarizes the error correction we have
used and their parameters for some useful architecture metrics.

\begin{table}
    \begin{center}
    \begin{footnotesize}
    \begin{tabular}{|l|l|l|}
    \hline
    \multicolumn{3}{|c|}{Error Correction Metric Summary} \\
    \hline
    Architecture Metric & Error Code - Level & Value \\
    \hline\hline
    EC Time (seconds)   &   \steane - L1        & 3.1 $\times 10^{-3}$    \\
                        &   \steane - L2        & 0.3                   \\
                        &   \bacon  - L1        & 1.2 $\times 10^{-3}$  \\
                        &   \bacon  - L2        & 0.1                   \\
    \hline\hline
    Qubit Size          &   \steane - L1        & 0.2                  \\
    (mm$^2$)            &   \steane - L2        & 3.4                  \\
                        &   \bacon  - L1        & 0.1                  \\
                        &   \bacon  - L2        & 2.4                  \\
    \hline\hline
    Transversal Gate    &   \steane - L1        & 6.2 $\times 10^{-3}$                \\
    Time (seconds)          &   \steane - L2        & 0.5                  \\
    &   \bacon  - L1        & 2.4 $\times 10^{-3}$ \\
                        &   \bacon  - L2        & 0.2                  \\
    \hline\hline
    Size, number of	&   \steane - L1          & 7 \\
    logical qubits	&   \steane - L1(ancilla) & 21\\
			&   \steane - L2	  & 49\\
			&   \steane - L2(ancilla) & 441\\
			&   \bacon - L1		  & 9\\
			&   \bacon - L1(ancilla)  & 12 \\
			&   \bacon - L2		  & 81\\
			&   \bacon - L2(ancilla)  & 298\\
    \hline
    \end{tabular}
    \end{footnotesize}
    \end{center}
    \caption{\textsf{Error Correction Metric Summary.
    Given the fact that we use optimistic ion-trap parameters all numbers are
    estimates and are thus rounded to only one significant digit.}}
    \label{table:code_metrics}
\end{table}

\subsection{Code Transfer Networks: Overview}
\begin{figure}[htbp]
\centering{
\includegraphics[width=3in]{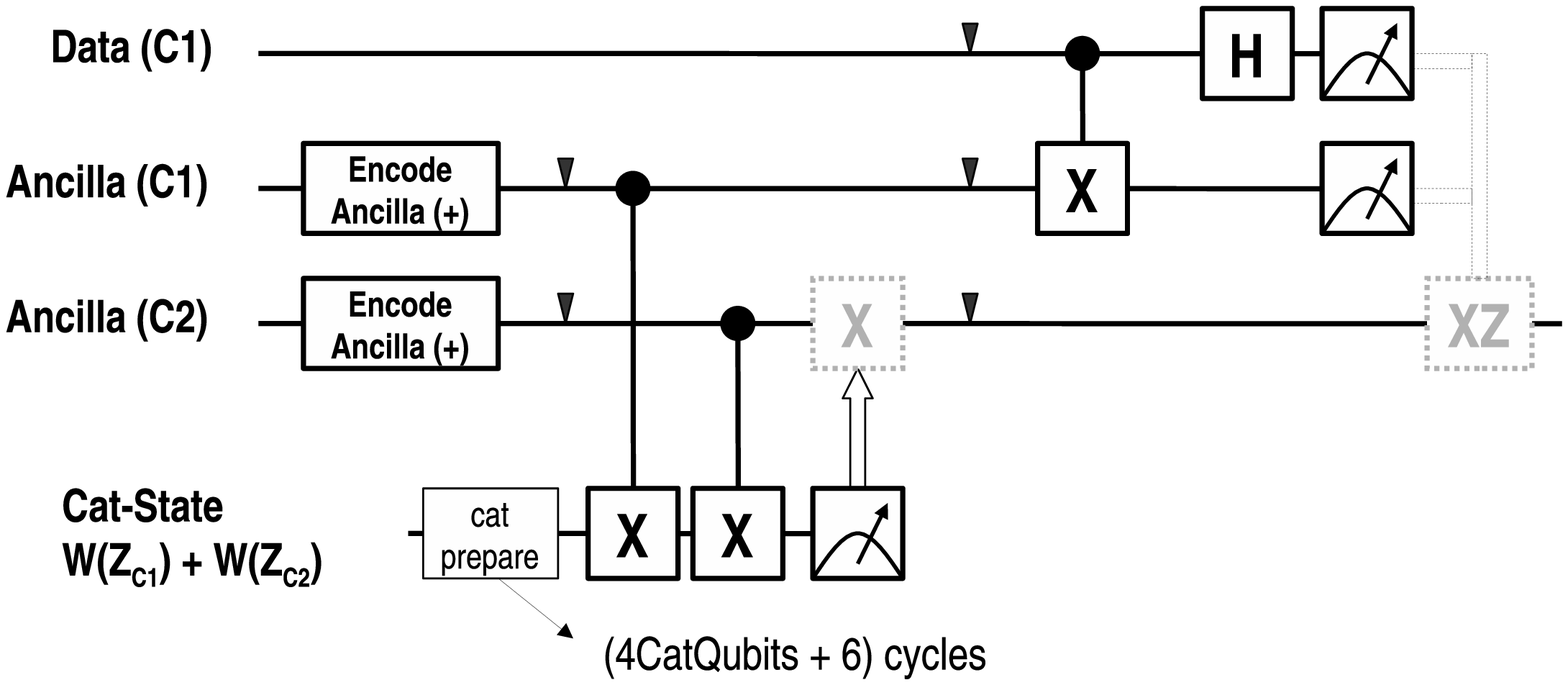}
\caption{\textsf{Code Teleportation Network from Code 1 (C1) to Code 2 (C2)
         C1 and C2 can even be the same error correcting code, but different
         levels of encoding. The solid triangles denote an error correction step.
         }}}\label{fig:transfer_network}
\end{figure}

\vspace{-0.5cm}
One of the most interesting components of the memory hierarchy are
the code transfer regions. This region transfers data encoded in
code C1 to a second code C2 without the need to decode. Figure
\ref{fig:transfer_network} illustrates this concept. The transfer
network {\em teleports} the data in C1 to C2, where C1 and C2 may
be any two error correcting codes. The code teleportation
procedure works much the same way as standard data teleportation
that is used for communication. A correlated ancillary pair is
prepared first between C1 and C2 through the use of a multi-qubit
cat-state (i.e. ``$(00...0 + 11...1)$''). The data qubit interacts
with the equivalently encoded ancillary qubit through a \cnot gate,
and the two are measured. Following the measurement the state of the
data is recreated at the C2 encoded ancillary qubit. This process is
required every time we transfer a qubit from memory to the cache or
vise-versa. Table \ref{table:transfer_times} summarizes the times 
for different code transfer combinations between levels 1 and 2 for
the \steane and the \bacon codes. 

\begin{table}
    \begin{center}
    \begin{footnotesize}
    \begin{tabular}{|l|l|l|l|l|}
        \hline
        (seconds)   &  7-L1     &   7-L2    & 9-L1      & 9-L2 \\
        \hline 
        7-L1        &     0     &   0.6    &   0.02    &  0.2  \\
        7-L2        &   1.3    &     0     &   1.3    &  1.5  \\
        9-L1        &   0.01    &   0.5    &      0    &  0.1  \\
        9-L2        &   0.4    &    0.9    &   0.4    &    0   \\
        \hline
    \end{tabular}
    \end{footnotesize}
    \end{center}
    \caption{\textsf{Transfer network latency for a combination of the
    \steane and \bacon codes.}}
    \label{table:transfer_times}
\end{table}

\section{CQLA Analysis and Results}\label{sec:cqla_optim}
\vspace{-0.5cm}

This section provides analysis of the abstractions presented in

to perform quantum modular exponentiation.

\subsection{Specialization into Memory}
\vspace{-0.5cm}
We now analyze our design, the CQLA, when it separates the 
quantum processor into
memory and compute regions. High density in memory is achieved by
greatly reducing the ratio of logical data qubits to logical ancilla qubits,
which is $(8:1)$ in memory and is $(1:2)$ in the compute regions.
This greatly reduces overall area
since prior work had a ratio of $(1:2)$ throughout the architecture.
Thus the memory is denser, but slower, which is permissible due to 
the large memory wait times \ref{table:params}. 

\begin{table*}[htb]
   \begin{center}
   \begin{tabular}{|c|c|c|c|c|c|c|c|}
   \hline
    Input & Compute & \multicolumn{2}{|c|}{Area Reduced (Factor of)} & \multicolumn{2}{|c|}{ SpeedUp }& \multicolumn{2}{|c|} {Gain Product}\\ \cline{3-8}
    Size&  Blocks& St-Code & BSr-Code & St-Code & BSr-Code & St-Code & BSr-Code\\ 
   \hline
   32-bit &  4&    6.69& 9.80  & 0.54 & 1.47 & 3.61 & 14.41\\
          &  9&    3.22& 4.74  & 0.97 & 2.9 & 3.14 & 13.74\\
   \hline
   64-bit &  9 &   6.36& 9.32 & 0.70 & 1.92 & 4.45 & 17.70 \\
        &    16&   3.79& 5.56  & 0.98 & 3.0 & 3.71 & 16.68 \\
   \hline
   128-bit&  16&   7.24&  10.6  & 0.72 & 1.97 & 5.24 & 20.88 \\
     &       25&   4.90&  7.17  & 0.96 & 2.84 & 4.70 & 20.36 \\
   \hline
   256-bit&  36&   6.65&  9.47  & 0.92 & 2.51 & 6.12 & 23.68 \\
     &       49&   5.07&  7.43  & 0.98 & 2.98 & 4.96 & 22.14 \\
   \hline
   512-bit&  64&   7.42&10.87  & 0.92 & 2.50 & 6.80 & 27.18 \\
     &       81&   6.06& 8.87  & 0.98 & 2.91 & 5.94 & 25.81 \\
   \hline
   1024-bit& \bf{100}&  \bf{9.14}& \bf{13.4}  & \bf{0.80} & \bf{2.19} & \bf{7.35} & \bf{29.35} \\
      &      121&  7.81&  11.45 & 0.97  & 2.65 & 7.60 & 30.34 \\
   \hline
   \end{tabular}
   \label{tab:area_result}
   \caption{\textsf{For various size inputs, this table shows how the CQLA
   performs for Modular Exponentiation. The space saved due to compressing
   the memory blocks and separating memory and compute regions is shown as
   compared to prior work \cite{Metodi05}. St-Code is the Steane ECC and 
   BSr-Code is the Bacon-Shor code. The Gain Product is compared
   with our prior work, the QLA, which has a Gain Product of 1.0.}}
\end{center}
\end{table*}

\begin{figure*}
\centering \subfigure[]{
    \label{fig:util}
    \includegraphics[width=3.0 in]{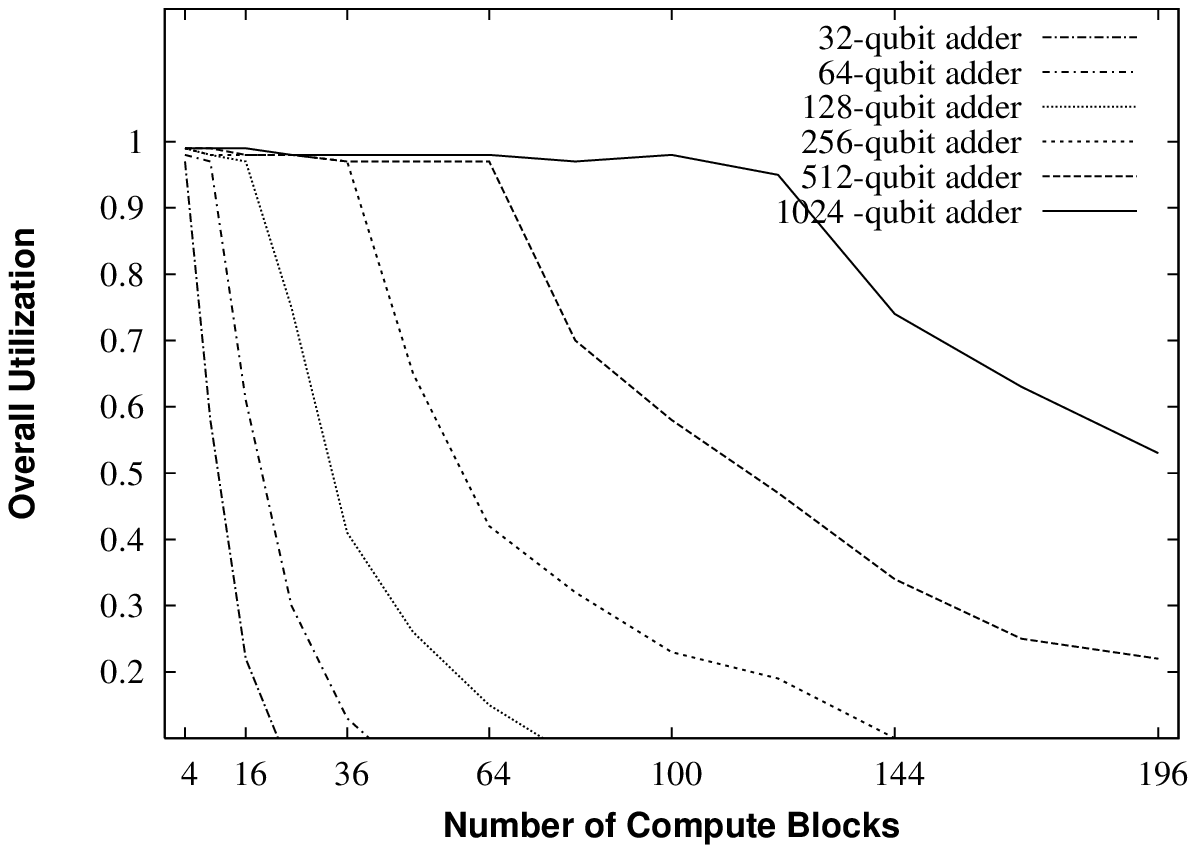}}
\hspace{1cm}\centering \subfigure[]{
    \label{fig:bandwidth}
    \includegraphics[width=3.0 in]{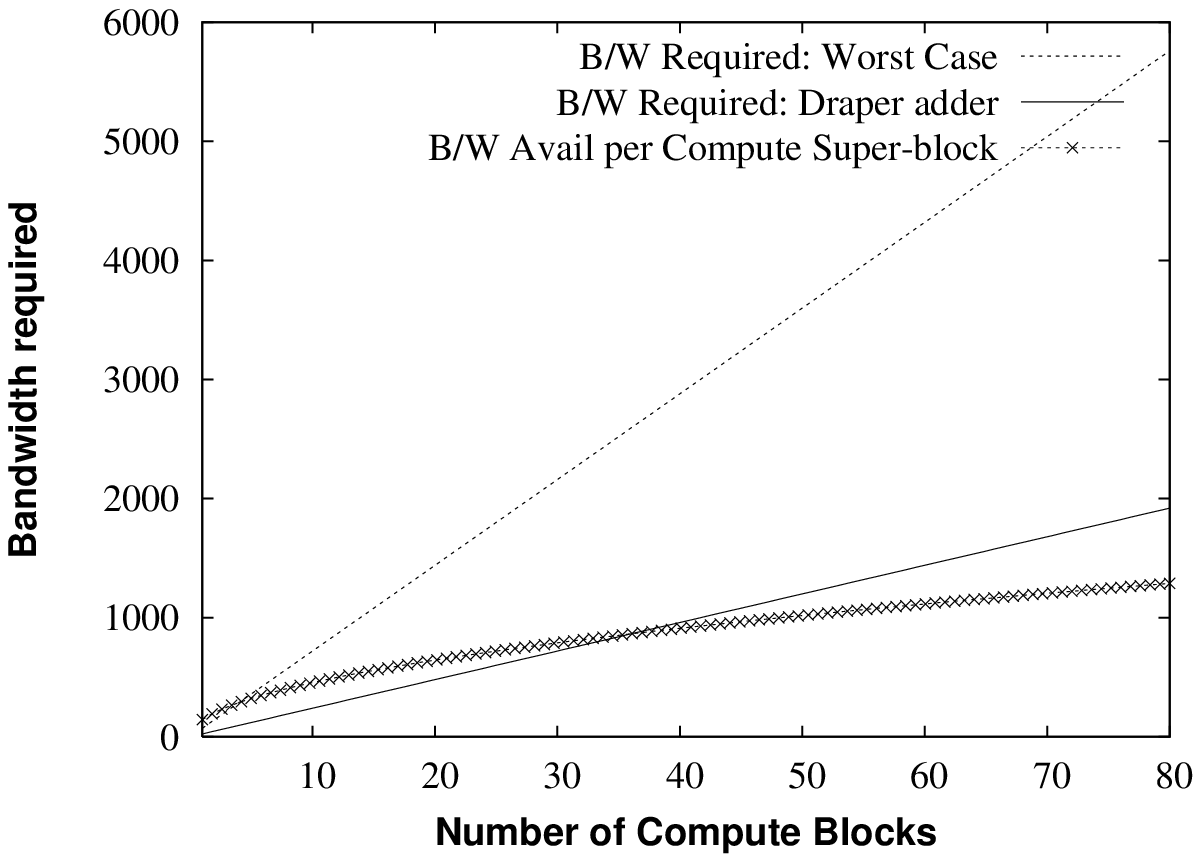}}
\caption{\textsf{{\bf (a)} 
Change in utilization as the number of compute blocks increases. 
{\bf (b)} 
The point of intersection of the two bottom curves is the optimal size of a compute superblock. These two curves are
bandwidth required (at the perimeter of the compute superblock) in
modular exponentiation and bandwidth available. The third steep curve, is
the worst case bandwidth required.}}
\label{fig:arch_util}
\end{figure*}

Quantum modular exponentiation is the most time consuming
part of Shor's algorithm, and the Draper carry-lookahead adder is its
most efficient implementation. This adder comprises single qubit gates,
two qubit cnot gates and three qubit toffoli gates and is dominated by
toffoli gates. The time to perform a single  fault-tolerant 
toffoli is equal to the time for fifteen two qubit gates, each of which
is followed by an error-correction step.
Table \ref{tab:area_result} shows the savings that can be achieved when
using denser memory. Note that performance is minimally impacted for 
the Steane Code as we exploit the limited parallelism in the adder.
We address the parallelism available within the application itself and
determine the number of compute blocks to maximally exploit this
parallelism with change with problem size. Figure \ref{fig:util}
shows how for a fixed problem size, utilization of each
compute block decreases with an increase the number of compute blocks.
Clearly, the decrease in utilization is offset by the increase in
overall performance. Thus the challenge here is to find the {\it
balance} between utilization and performance.

We compare all our results to \cite{Metodi05}, which used only the Steane ECC.
Since the Bacon-Shor ECC uses fewer overall resources \ref{table:code_metrics} and 
allows faster error-correction, a design based on these codes not only is
much smaller, but is also faster. 
The CQLA, thus reduces area required by a {\bf factor of 9} with minimal
 performance reduction for the Steane ECC and by a {\bf factor of 13} 
 with a {\bf speedup of 2} when using the Bacon-Shor ECC.
To compare the relative merit our design choices, we use the {\em gain product}
which can be defined by $GP = \left(Area_{old} * AdderTime_{old}\right) /
\left(Area_{CQLA} * AdderTime_{CQLA}\right)$ where $AdderTime$ is
the average time per adder for modular exponentiation. The gain product
indicates the improvement in system parameters relative to our 
prior work, the QLA.  The higher the gain product, the better the collective
improvement in area and time of our system.

{\bf Communication Issues: }Toffoli gates cannot be directly implemented 
on encoded data and have to be broken down into 
multiple two qubit gates. Performing
a fault-tolerant Toffoli between three logical qubits requires extra
logical ancilla and logical cat-state qubits. The 
flow of data between these nine qubits to complete a single
toffoli forms the most intense communication pattern during the entire addition operation.
To study the bandwidth requirements during the toffoli gates, we 
developed a scheduler that would try to have all the requirements for
communication (creating EPR pairs, transporting and purifying them) in place
while the logical qubit to be transported was undergoing error-correction
after completion of the previous gate.
With bandwidth of one channel, it was possible to overlap communication
with computation for the Steane $[[7,1,3]]$ code. To enable this overlap
when using the Bacon-Shor code, the required bandwidth was three channels.
Table \ref{table:code_metrics}
shows that while a logical qubit encoded in the Bacon-Shor code is smaller 
when ancilla are considered; it has more data qubits than the Steane code.
Since only data qubits are involved during teleportation, the time
for teleporting a logical qubit in the Bacon-Shor code is greater. In addition,
the Bacon-Shor codes take far fewer error-correction cycles. These two
factors push its bandwidth requirement higher. Note that the higher bandwidth
is accounted for in results of Table \ref{tab:area_result}.

{\bf Superblocks:} In the CQLA, several compute blocks together
form compute superblocks. This is done to exploit the locality
inherent to an application. Having larger superblocks also increases
the perimeter bandwidth between the compute and memory regions
of the CQLA. This increase in bandwidth of a larger superblock
is offset by the much greater increase in communication required.
Our intuition tells us that at a certain point, it may be more
efficient to have multiple small superblocks instead of one large
superblock. To determine this number concretely, we plot the
change in bandwidth required against change in bandwidth available.
Figure \ref{fig:bandwidth} shows the cross-over point
is 36 compute blocks per superblock, immaterial of what error correction
code is used. Thereafter it is no longer beneficial
to increase the size of an individual compute superblock.

\subsection{Memory Hierarchy}
\vspace{-0.5cm}
Reducing the encoding level of the compute region will dramatically
increase its speed. Recall that resources, time and reliability
all increase exponentially as we increase the level of encoding. With
the compute region at level 1 and memory at level 2, the challenge is
the very familiar one of the CPU being an order of magnitude faster 
than the memory. To maximize the benefit of a much faster compute
region, we introduce the quatum memory hierarchy. In our hierarchy,
the memory is at level 2 encoding (slow and reliable), cache is at level 1 
(faster, less reliable) and the compute region is also at level 1 (fastest
and same reliability as cache). The difference in speed between the compute
region and the cache is the due to a greater number of ancilla in the
compute region.

To study the behavior of the CQLA with a cache and multiple encoding levels,
we developed a simulator that models a cache. The simulator takes into
account the computation cost in both encoding levels and also the cost
of transferring logical qubits between encoding levels. The application under
consideration is still the Draper carry-lookahead adder. Input to the simulator 
is a sequence of instructions; each instruction is similar to assembly language
and describes a logical gate between qubits. We have written generators
that output this code in a form that can take advantage of an architecture
with maximal parallelism.

\begin{figure}[htbp]
\centering{
\includegraphics[width=3.0 in]{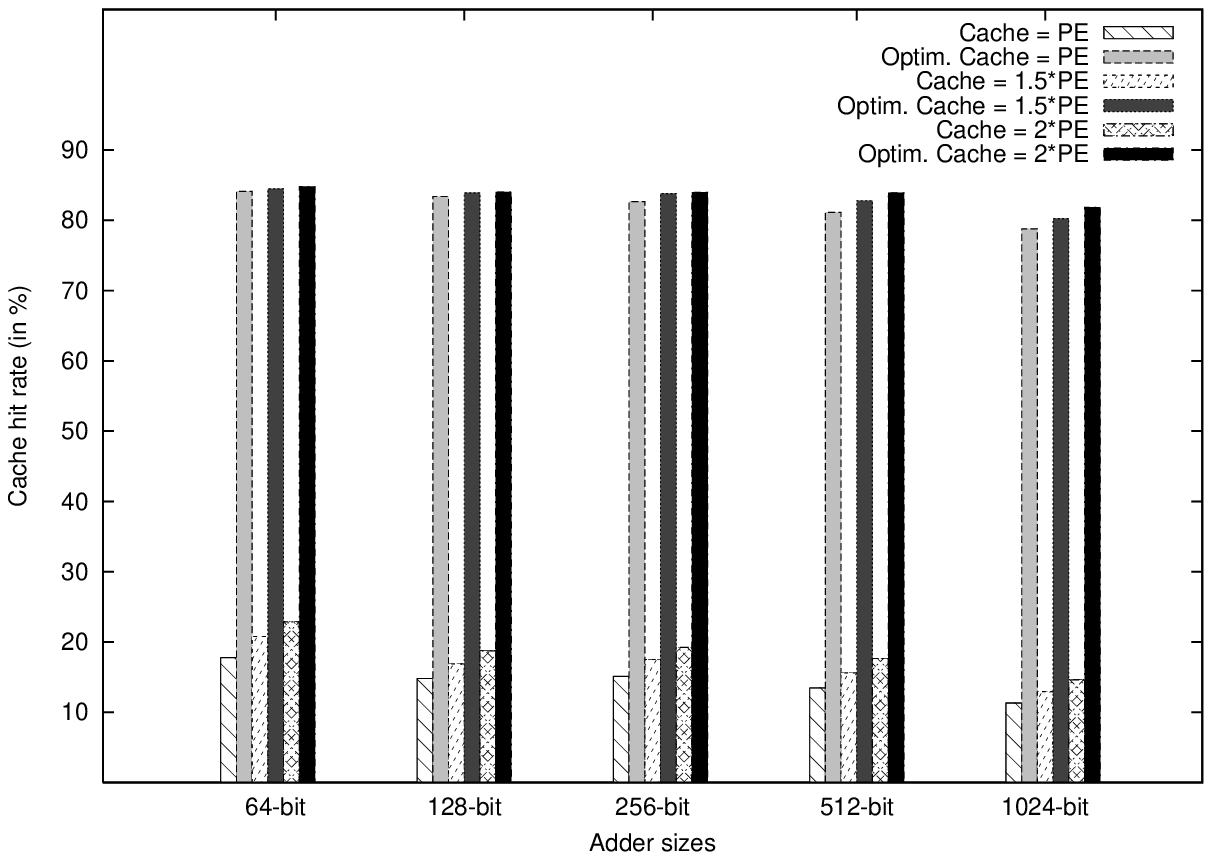}}
\caption{\textsf{ Shows the cache hitrate for different adders 
when both cache and 
compute region are at Level 1 recursion. Largest cache considered
holds twice the number of logical qubits as the compute block. Results for
both the non-optimized version and the optimized version are shown.}}
\label{fig:cache_hitrate}
\end{figure}

When the simulator runs this code in the sequence intended by the Draper 
carry-lookahead adder, the cache hit-rate is limited to
$20\%$. To improve the hit-rate, we utilize the following optimized approach. 
Since 
we are scheduling statically, the instruction fetch window for the simulator
can be the whole program. The simulator takes advantage of this by first
creating a dependency list of all input instructions. Then it carefully 
selects the next instruction such that probability of finding all required 
operands in the cache is maximized. 
This optimized fetch yields a cache hit-rate of 
almost $85\%$ immaterial of adder size and cache size.
The replacement policy in the cache is least recently used. 
Figure \ref{fig:cache_hitrate} shows the cache hit-rates for
different sized adders for the non-optimized and optimized instruction
fetch approaches. If $n$ is the number of logical qubits in the compute region,
the cache sizes we studied were $n, 1.5n$ and $2n$.
As the graph shows, the increase in hit-rate is more 
pronounced due to the optimized fetch than increasing cache size. For the
CQLA, we thus employ a cache size of twice the number of qubits in the
compute region. The high hit-rate means the transfer networks will not be
overwhelmed. 

\begin{table*}
   \begin{center}
   \begin{tabular}{|c|c|c|c|c|c|c|}
   \hline
   Par Xfer & Adder Size & L1 SpeedUp & L2 SpeedUp & Adder SpeedUp & Area Reduced & Gain Product \\
   \hline
   \multicolumn{7}{|c|}{Steane $[[7,1,3]]$ Code} \\
   \hline
      & 256  &17.417 & 0.98 & 6.25 & 5.07 & 31.68 \\
   10 & 512  & 17.41 & 0.97 & 6.33 & 6.06 & 38.38 \\
      & 1024 & 18.18 & 0.88 & \bf{4.93} & \bf{9.14} & \bf{45.06} \\
   \hline
      & 256 & 10.409 & 0.98 & 4.05 & 5.07 & 24.99 \\
   5  & 512 & 10.408 & 0.97 & 4.04 & 6.06 & 24.48 \\
      &1024 & 10.96  & 0.88 & \bf{2.94} & \bf{9.14} & \bf{26.87} \\
   \hline
   \multicolumn{7}{|c|}{Bacon-Shor $[[9,1,3]]$ Code} \\
   \hline
      & 256  & 9.61 & 1.53  & 5.92 & 7.43 & 43.99 \\
   10 & 512  & 9.61 & 2.28  & 8.82 & 8.87 & 78.23 \\
      & 1024 & 10.15& 2.00  & \bf{8.10} & \bf{13.4} & \bf{108.53} \\
   \hline
      & 256 & 5.17  & 1.53  & 3.66 & 7.43 & 27.19 \\
   5  & 512 & 5.17  & 2.28  & 5.45 & 8.87 & 48.37 \\
      &1024 & 5.49  & 2.00  & \bf{4.99} & \bf{13.40}& \bf{66.90} \\
   \hline
   \end{tabular}
   \caption{\textsf{This table shows the results of incorporating a 
   memory hierarchy and two separate encoding levels. Depending on the
   number of parallel transfers possible between memory and cache, we
   can expect different speedup values for the adder at level 1. This 
   combined with results from Table \ref{tab:area_result} give us the
   final Gain Product. Comparatively, prior work has an Gain Product
   number of 1.0. }}
   \label{tab:time_result}
\end{center}
\end{table*}

\noindent{{\bf Fault-tolerance with multiple encoding levels: }
A quantum computer running an application of size $S = KQ$, where
$K$ is the number of time-steps and $Q$ is the number of logical
qubits, will need to have a component failure rate of at most $P_f =
1/KQ$. To evaluate the expected component failure rate at some level
or recursion we use Gottesman's estimate for local architectures
\cite{Gottesman99} shown in Equation~\ref{eqn:ftrecurselocal} below.

\begin{equation}
    P_f = \frac{1}{cr^2r^L}(cr^2p_0)^{2^L} =
         \frac{p_{th}}{r^L}(p_{th}^{-1}p_0)^{2^L}
    \label{eqn:ftrecurselocal}
\end{equation}

The value for $r$ is the communication distance between level $1$
blocks which are aligned in QLA to allow $r=12$ cells on average and
$L$ denotes the level of recursion. The threshold failure rate,
$p_{th}$, for the Steane $[[7,1,3]]$ circuit accounting for movement and
gates was computed in \cite{Svore04a} to be approximately $7.5
\times 10^{-5}$. Taking as $p_0$ the average of the expected failure
probabilities given in Table \ref{table:params}, and using
Equation \ref{eqn:ftrecurselocal}, we find that for our system to 
be reliable it can spend {\bf only 2\%} of the total execution time in level 1.
Recall that error-correction is the most frequently peformed operation
in the CQLA. For the Steane code, level 2 error correction takes 0.3 sec
and level 1 takes $3.1 \times10^{-3}$ sec, which is approximately  $1\%$ of the 
level 2 time.
Thus if all operations performed by the CQLA were equally divided between
level 1 and level 2 operations, the system will maintain its fidelity. 
The Bacon-Shor ECC can be analyzed in a similar manner and their results
are more favourable due to a higher threshold.

The CQLA architecure now consists of a memory at level 2, a compute region 
also at level 2, a cache and a compute region at level 1 and transfer networks
for changing the qubit encoding levels. Since quantum modular exponentiation 
is perfomed by repeated quantum additions, we
could perform half of these additions completely in level 2 and the other half
in level 1. To comfortably maintain the fidelity of the system, we perform
one level 1 addition for every two level 2 additions. The resulting 
increase in performance is shown in Table \ref{tab:time_result}.

\section{Application Behavior}\label{sec:app_results}
\vspace{-0.5cm}

\begin{figure*}
\centering \subfigure[]{
    \label{fig:shor_a}
    \includegraphics[width=3.0 in]{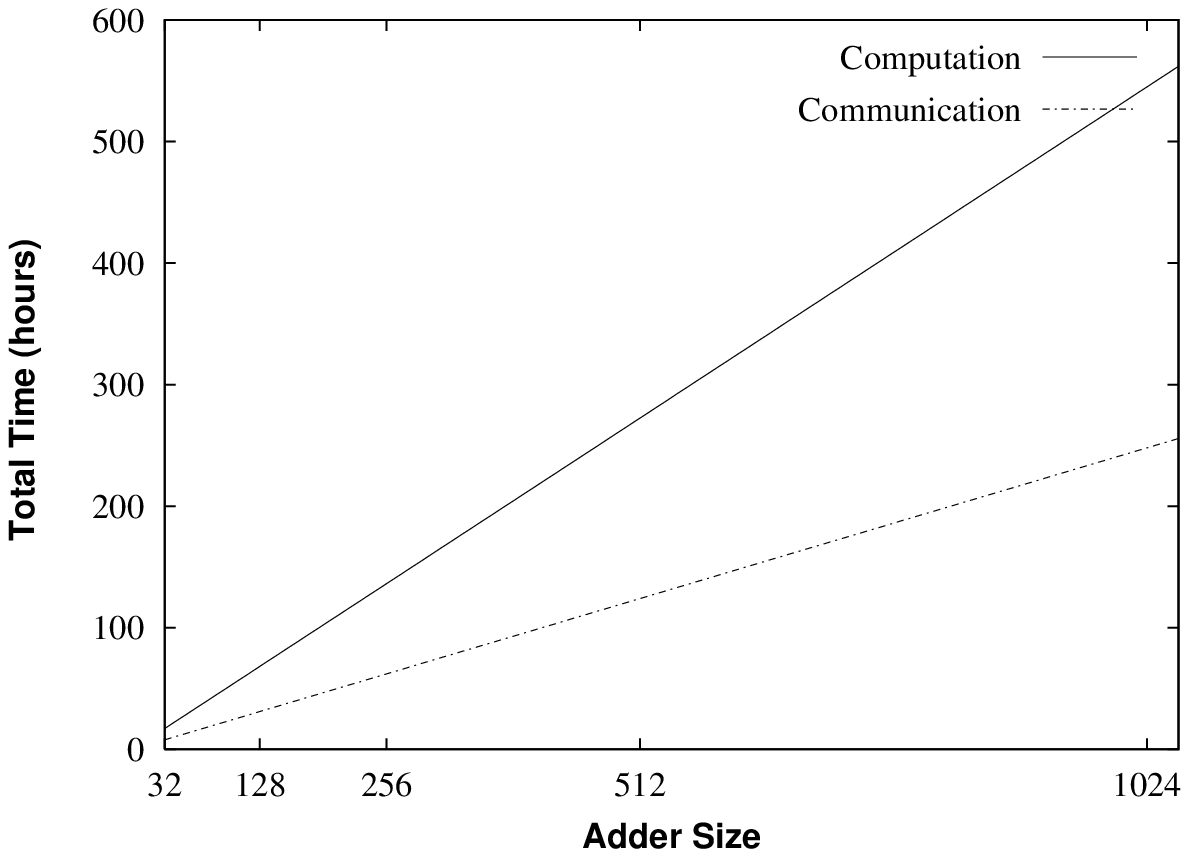}}
\hspace{0.5cm}\centering \subfigure[]{
    \label{fig:shor_b}
    \includegraphics[width=3.0 in]{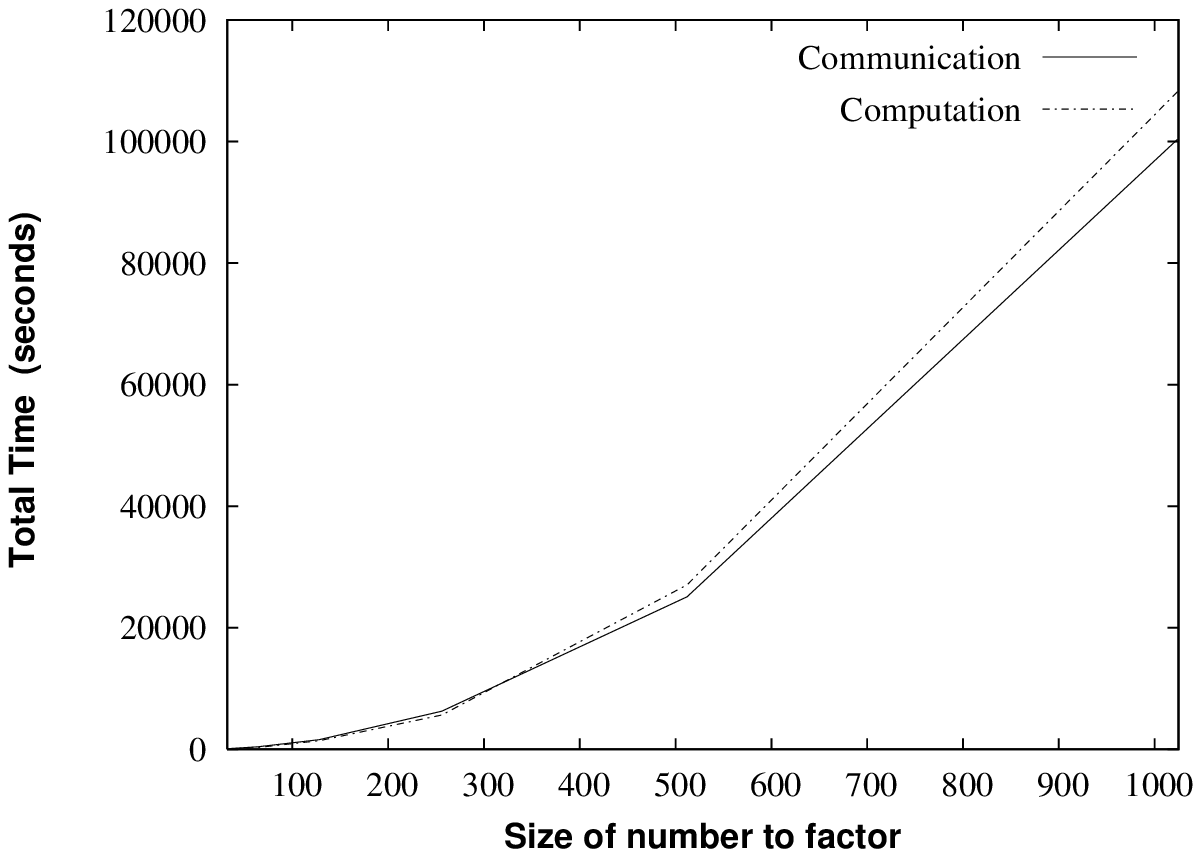}}
\caption{\textsf{Total communication and computation times for the two
components of Shor's algorithm, {\bf (a)} Modular Exponentiation {\bf
(b)} Quantum Fourier Transform (QFT). Although communication is
significant in the QFT, Modular exponentiation dominates Shor's
algorithm. Both these results are for the Bacon-Shor code}}
\label{fig:apps}
\end{figure*}

In this secti 
compute and memory are at level 2 encoding.
Contrary to traditional silicon based processors, in
the CQLA a single communication step does not take longer than the
computation of a single gate. The reason behind this phenomenon is
the lack of reliability of quantum data, which forces us to perform
an error-correction procedure after each gate. 
The time to complete a
fault-tolerant Toffoli is about 20 times greater than a two-qubit
\cnot gate. The applications we study are modular exponentiation 
and the quantum fourier transform.

\subsection{Shor's Algorithm}
\vspace{-0.5cm}

Shor's algorithm is the most celebrated of quantum algorithms due to
its potential exponential advantage over conventional algorithms and
its application to breaking public-key cryptography \cite{Shor94}.
Shor's algorithm is primarily composed of two parts, the modular
exponentiation and the quantum fourier transform.

\noindent {\bf Modular Exponentiation: } The execution of 
modular exponentiation is dominated by Toffoli gates.
To keep the compute block from having to wait for
qubits, and hence stalling, the bandwidth around the perimeter of
the compute block has to accommodate the transfer of three qubits to
and from memory. Intuitively, since the CQLA is a mesh, and the
bottleneck in bandwidth will be at the edge of the compute blocks,
having adequate bandwidth at this edge is sufficient for the rest of
the mesh.

Based on the communication results from \cite{Metodi05}, we
calculate that a $2$ channels on the perimeter of the compute block
would provide adequate bandwidth for all required communication. 
We compute the time required for all communication
steps and compare it against the total computation time for
differently sized adders. The result is shown in \ref{fig:shor_a}
and demonstrates that communication requirements do not adversely
impact the design.

\noindent {\bf Quantum Fourier Transform: }
While the Quantum Fourier Transform (QFT) comprises a small fraction of
the overall Shor's algorithm, it requires all-to-all personalized
communication between data qubits. In addition, it uses only
one-qubit and two-qubit gates which consume much less time. As a
result, studying the performance of the QFT gives us an insight into
how the CQLA will behave when faced with an communication heavy and
a computation light application.

In the worst case, all nine data qubits (maximum capacity of the
compute block) would have to be transferred to or from memory
simultaneously.

Between compute blocks, the QFT's all-to-all personalized
communication must be supported on the CQLA mesh network.  We leverage
the vast amount of prior work done in studying mesh networks, and
employ a near-optimal algorithm proposed in \cite{Yang04}. The  total time
for communication for varying problem sizes is shown in figure
\ref{fig:shor_b}.  Note that while communication time is a little
less, it closely tracks the computation time for all problem
sizes. This is due to the difference in time to error correct a single
logical qubit and the time to transport a single qubit; which stays
constant immaterial of the problem size.

\section{Future Work}\label{sec:issues}

A high-level goal of this work is to build abstractions from which
architects and systems designers can examine open issues and help
guide the substantial basic science and engineering under investment
towards building a scalable quantum computer.  
The primary focus of our work has been system balance.  The driving
force in this balance has been application parallelism.  A key open
issue is the restructuring of quantum algorithms to manage this
parallelism in the context of system balance.  From an architectural
point of view, the most relevant abstract properties are density of
functional components, the memory hierarchy  and communication bandwidth.  

While our work
has focused on trapped ions, most scalable technologies will have a
similar two-dimensional layout where our techniques can be easily applied.
This is because the density is determined by the ratio of data to ancilla
rather than physical details of the underlying technology.

For ion-traps, lasers can also be a control issue. We plan to study
how our architecture can minimize the number of lasers and minimize
the power consumed by each laser, since power is proportional to
fanout. Efficiently routing control signals to all electrodes in an ion-trap
is a challenging proposition, one that has not yet been considered for large
systems. Currently, we perform the whole adder at the fast level $1$ encoding
or at the level $2$ encoding; clever instruction scheduling techniques
can allow us to improve performance by reducing granularity.

\section{Conclusion}\label{sec:conclude}
The technologies and abstractions for quantum computing have evolved
to an exciting stage, where architects and system designers can
attack open problems without intimate knowledge of the physics of
quantum devices.  We explore the amount of parallelism available in
quantum algorithms and find that a specialized architecture can
serve our needs very well. The CQLA design is an example where architectural
techniques of specialization and balanced system design have led to
up to a {\bf 13X} improvement in density and a {\bf 8X} increase in 
performance, while preserving fault tolerance. 
We hope that further application of compiler and
system optimizations will lead to even more dramatic gains towards a
scalable, buildable quantum computer.

\singlespacing 
\begin{footnotesize}
\bibliographystyle{spiebib}
\bibliography{isca_refs}
\end{footnotesize}

\end{document}

%% file: macros.tex
\def\>{\rangle}
\def\be{\begin{equation}}
\def\ee{\end{equation}}

\newcommand{\gt}[1]{\textsc{#1}}
\newcommand{\cnot}{\gt{cnot~}}

\newcommand{\eccode}[3]{$[\![#1,#2,#3]\!]\;$}
\newcommand{\steane}{\eccode{7}{1}{3}}
\newcommand{\bacon}{\eccode{9}{1}{3}}
\newcommand{\ecc}{\eccode{n}{k}{d}}

\renewcommand{\em}{\it}

\newcommand{\comment}[1]{}
\newcommand{\ignore}[1]{}

\def\ebcfigure[#1,#2,#3]{
\begin{figure}
\vspace*{4mm}
\begin{center}

\epsfxsize=3in\
\hspace*{-0.5in}\
\vspace*{6mm}\
\epsfbox{#1}\

\vspace*{0mm}\caption[]{#2
} \label{#3}

\vspace*{-5mm}
\end{center}
\vspace*{4mm}
\end{figure}}
\def\bcfigure[#1,#2,#3]{
\begin{figure}
\vspace*{4mm}
\begin{center}

\epsfxsize=2in\
\epsfbox{#1}\

\vspace*{-3mm}\caption[]{#2
} \label{#3}

\vspace*{-5mm}
\end{center}
\vspace*{4mm}
\end{figure}}

\def\bdcfigure[#1,#2,#3,#4,#5,#6]{
{
\begin{figure*}
\vspace*{0.2in}\
\begin{center}
\begin{minipage}{3in}{
\epsfxsize=3in\
\epsfbox{#1}
\vspace*{-3mm}\caption[]{#2} \label{#3} \
}\end{minipage}\hspace*{0.5in}\
\begin{minipage}{3in}{
\epsfxsize=3in\
\epsfbox{#4}
\vspace*{-3mm}\caption[]{#5}\label{#6} \
}\end{minipage}
\end{center}
\vspace*{-0.4in}\
\end{figure*}
}
}

\def\cfigure[#1,#2,#3]{
\begin{figure}
\vspace*{4mm}
\begin{center}

\epsfxsize=2in\
\epsfbox{#1}\

\vspace*{-3mm}\caption[]{#2
} \label{#3}

\vspace*{-5mm}
\end{center}
\vspace*{4mm}
\end{figure}}

\def\dcfigure[#1,#2,#3,#4,#5,#6]{
{
\begin{figure*}
\begin{center}
\vspace{0.5in}\
\begin{minipage}{3in}{
\epsfxsize=2.0in\
\epsfbox{#1}
\vspace*{-3mm}\caption[]{#2} \label{#3} \
}\end{minipage}\hspace*{0.5in}\
\begin{minipage}{3in}{
\epsfxsize=2.0in\
\epsfbox{#4}
\vspace*{-3mm}\caption[]{#5}\label{#6} \
}\end{minipage}
\end{center}
\vspace*{-0.4in}\
\end{figure*}
}
}

\def\t3cfigure[#1,#2,#3]{
{
\begin{figure*}
\hspace*{-0.2in}\
\vspace*{0.2in}\
\begin{center}
\hspace*{-0.3in}\
\begin{minipage}{3in}{
\epsfxsize=2in\
\epsfbox{vortex/#1}
\vspace*{0mm}\\
{(a) Vortex}
}\end{minipage}\hspace*{-0.5in}\
\begin{minipage}{3in}{
\epsfxsize=2in\
\epsfbox{perl/#1}
\vspace*{0mm}\\
{(b) Perl}
}\end{minipage}\hspace*{-0.5in}\
\begin{minipage}{3in}{
\epsfxsize=2in\
\epsfbox{li/#1}
\vspace*{0mm}\\
{(c) Xlisp}
}\end{minipage}
\end{center}
\vspace*{0.0in}\caption[]{#2}\label{#3}\
\end{figure*}
}
}

\def\st3cfigure[#1,#2,#3]{
{
\begin{figure*}
\hspace*{-0.2in}\
\vspace*{0.2in}\
\begin{center}
\hspace*{-0.3in}\
\begin{minipage}{3in}{
\epsfxsize=1.5in\
\epsfbox{vortex/#1}
\vspace*{0mm}\\
{(a) Vortex}
}\end{minipage}\hspace*{-0.5in}\
\begin{minipage}{3in}{
\epsfxsize=1.5in\
\epsfbox{perl/#1}
\vspace*{0mm}\\
{(b) Perl}
}\end{minipage}\hspace*{-0.5in}\
\begin{minipage}{3in}{
\epsfxsize=1.5in\
\epsfbox{li/#1}
\vspace*{0mm}\\
{(c) Xlisp}
}\vspace*{0.25in}\end{minipage}
\end{center}
\vspace*{0.0in}\caption[]{#2}\label{#3}\
\end{figure*}
}
}

\def\vt3cfigure[#1,#2,#3]{
{
\begin{figure*}
\hspace*{-0.2in}\
\vspace*{0.2in}\
\begin{center}
\hspace*{-0.3in}\
\begin{minipage}{3in}{
\epsfxsize=3.5in\
\epsfbox{vortex/#1} \\
{(a) Vortex}
\vspace*{0.5in}
}\end{minipage} \\
\hspace*{-0.5in}\
\begin{minipage}{3in}{
\epsfxsize=3.5in\
\epsfbox{perl/#1} \\
{(b) Perl}
\vspace*{0.5in}
}\end{minipage} \\
\hspace*{-0.5in}\
\begin{minipage}{3in}{
\epsfxsize=3.5in\
\epsfbox{li/#1} \\
{(c) Xlisp}
\vspace*{0.5in}
}\end{minipage}
\end{center}
\vspace*{0.0in}\caption[]{#2}\label{#3}\
\end{figure*}
}
}

\def\quadfigure[#1,#2,#3,#4,#5,#6]{
{
\begin{figure*}
\vspace*{0.2in}\
\begin{center}
\begin{minipage}{3in}{
\epsfysize=1.0in\
\epsfbox{#1}
}\end{minipage}\hspace*{0.5in}\
\begin{minipage}{3in}{
\epsfysize=1.0in\
\epsfbox{#2}
}\end{minipage}
\end{center}
\vspace*{-0.2in}\
\begin{center}
\begin{minipage}{3in}{
\epsfysize=1.0in\
\epsfbox{#3}
}\end{minipage}\hspace*{0.5in}\
\begin{minipage}{3in}{
\epsfysize=1in\
\epsfbox{#4}
}\end{minipage}
\end{center}
\vspace*{-0.3in}\
\caption{#5}\label{#6}
\end{figure*}
}
}

\def\wcfigure[#1,#2,#3]{
\begin{figure*}
\vspace*{4mm}
\begin{center}

\epsfxsize=5in\
\epsfbox{#1}\

\vspace*{-3mm}\caption[]{#2
} \label{#3}

\vspace*{-5mm}
\end{center}
\end{figure*}}

\def\wycfigure[#1,#2,#3]{
\begin{figure*}
\vspace*{4mm}
\begin{center}

\epsfysize=3.0in\
\epsfbox{#1}\

\vspace*{-3mm}\caption[]{#2
} \label{#3}

\vspace*{5mm}
\end{center}
\end{figure*}}

%% file: isca06-cqla.bbl
\begin{thebibliography}{10}

\bibitem{Metodi05}
T.~S. Metodi, D.~D. Thaker, A.~W. Cross, F.~T. Chong, and I.~L. Chuang, ``A
  quantum logic array microarchitecture: Scalable quantum data movement and
  computation,'' {\em Proceedings of the 38th International Symposium on
  Microarchitecture}~{\bf MICRO-38}, 2005.

\bibitem{Zurek}
W.~Wootters and W.~Zurek, ``A single quantum cannot be cloned,'' {\em
  Nature}~{\bf 299}, pp.~802--803, 1982.

\bibitem{Steane97a}
A.~M. Steane, ``Space, time, parallelism and noise requirements for reliable
  quantum computing,'' {\em Fortsch.~Phys.}~{\bf 46}, pp.~443--458, 1998.

\bibitem{Bacon05}
D.~Bacon, ``Operator quantum error correcting subsystems for self-correcting
  quantum memories,'' {\em quant-ph/0506023} , 2005.

\bibitem{Poulin05}
D.~Poulin, ``Stabilizer formalism for operator quantum error correction,'' {\em
  quant-ph/0508131} , 2005.

\bibitem{Aliferis05b}
P.~Aliferis {\em Unpublished work based on private conversations with Andrew
  Cross.} , 2006.

\bibitem{Oskin02}
M.~Oskin, F.~Chong, and I.~Chuang, ``A practical architecture for reliable
  quantum computers,'' {\em IEEE Computer} , January~2002.

\bibitem{Dur98a}
W.~Dur, H.~J. Briegel, J.~I. Cirac, and P.~Zoller, ``Quantum repeaters based on
  entanglement purification,'' {\em Phys. Rev.}~{\bf A59}, p.~169, 1999.

\bibitem{Steane96}
A.~Steane, ``Error correcting codes in quantum theory,'' {\em Phys. Rev.
  Lett}~{\bf 77}, pp.~793--797, 1996.

\bibitem{Cirac95}
J.~I. Cirac and P.~Zoller, ``Quantum computations with cold trapped ions,''
  {\em Phys. Rev. Lett}~{\bf 74}, pp.~4091--4094, 1995.

\bibitem{Sorensen00}
A.~Sorensen and K.~Molmer, ``Entanglement and quantum computation with ions in
  thermal motion,'' {\em Phys. Lett. A}~{\bf 62}, p.~02231, 2000.

\bibitem{Leibfried03}
D.~Leibfried and et~al., ``Experimental demonstration of a robust,
  high-fidelity geometric two ion-qubit phase gate,'' {\em Nature}~{\bf 422},
  pp.~412--415, 2003.

\bibitem{Hahn50}
E.~Hahn, ``Spin echoes,'' {\em Phys. Rev.}~{\bf 80}, pp.~580--594, 1950.

\bibitem{Wineland98}
D.~Wineland and et~al., ``Experimental issues in coherent quantum-state
  manipulation of trapped atomic ions,'' {\em Journal of Research of NIST}~{\bf
  103}, pp.~259--328, 1998.

\bibitem{Kielpinski02}
D.~Kielpinski, C.~Monroe, and D.~Wineland, ``Architecture for a large-scale
  ion-trap quantum computer,'' {\em Nature}~{\bf 417}, pp.~709--711, 2002.

\bibitem{Williams03a}
J.~Porto, S.~Rolston, T.~Laburthe, C.~Williams, and W.~Phillips, ``Quantum
  information with neutral atoms as qubits,'' {\em Phil. Trans. R. Soc.
  Lond.}~{\bf A361}, pp.~1417--1427, 2003.

\bibitem{Kielpinski99}
B.~Blinov, L.~Deslauriers, P.~Lee, M.~Madsen, R.~Miller, and C.~Monroe,
  ``Sympathetic cooling of trapped ions for quantum logic,'' {\em Phys. Rev.
  A.}~{\bf 61}, p.~032310, 2000.

\bibitem{Blinov02}
B.~Blinov, L.~Deslauriers, P.~Lee, M.~Madsen, R.~Miller, and C.~Monroe,
  ``Sympathetic cooling of trapped cd+ isotopes,'' {\em Phys. Rev. A.}~{\bf
  65}, p.~040304, 2002.

\bibitem{Barrett04}
M.~Barrett, J.~Chiaverini, T.~Schaetz, J.~Britton, and et. al., ``Deterministic
  quantum teleportation of atomic qubits,'' {\em Nature}~{\bf 429}, 2004.

\bibitem{Riebe04}
M.~Riebe, H.~Haffner, C.~Roos, and et. al., ``Deterministic quantum
  teleportation with atoms,'' {\em Nature}~{\bf 429}(6993), pp.~734--737, 2004.

\bibitem{Chiaverini05}
J.~Chiaverini, R.~B.~J. Britton, J.~Jost, C.~Langer, D.~Leibfried, R.~Ozeri,
  and D.~Wineland, ``Surface-electrode architecture for ion-trap quantum
  information processing,'' {\em E-Print: quant-ph/0501147} , 2004.

\bibitem{Kim05}
J.~Kim, S.~Pau, Z.~Ma, H.~McLellan, J.~Gages, A.~Kornblit, and R.~Slusher,
  ``System design for large-scale ion trap quantum information processor,''
  {\em Quantum Information and Computation}~{\bf 5(7)}, 2005.

\bibitem{ARDA}
D.~Wineland and T.~Heinrichs, ``Ion trap approaches to quantum information
  processing and quantum computing,'' {\em A Quantum Information Science and
  Technology Roadmap} , 2004.
\newblock URL: http://quist.lanl.gov.

\bibitem{Steane04b}
A.~Steane, ``How to build a 300 bit, 1 gop quantum computer,'' {\em
  arXiv:quant-ph/0412165} , 2004.

\bibitem{Ozeri05}
R.~Ozeri and et. al., ``Hyperfine coherence in the presence of spontaneous
  photon scattering,'' {\em arXiv:quant-ph/0502063} , 2004.

\bibitem{Wineland05}
D.~Wineland, D.~Leibfried, M.~Barrett, A.~Ben-Kish, and et.al., ``Quantum
  control, quantum information processing, and quantum-limited metrology with
  trapped ions,'' {\em Proceedings of the International Conference on Laser
  Spectroscopy (ICOLS)} , 2005.

\bibitem{Hensinger05}
W.~K. Hensinger, S.~Olmschenk, D.~Stick, D.~Hucul, M.~Yeo, M.~Acton,
  L.~Deslauriers, J.~Rabchuk, and C.~Monroe, ``T-junction ion trap array for
  two-dimensional ion shuttling, storage and manipulation,'' {\em E-Arxiv:
  quant-ph/0508097} , 2005.

\bibitem{Oskin05a}
S.~Balensiefer, L.~Kregor-Stickles, and M.~Oskin, ``An evaluation framework and
  instruction set architecture for ion-trap based quantum
  micro-architectures.,'' {\em ISCA-32; Madison, WI} , 2005.

\bibitem{Meter05}
R.~V. Meter and M.~Oskin, ``Architectural implications of quantum computing
  technologies,'' {\em ACM Journal on Emerging Technologies in Computing
  Systems (JETC)}~{\bf 2(1)}, 2006.

\bibitem{Draper04}
T.~Draper, S.~Kutin, E.~Rains, and K.~Svore, ``A logarithmic-depth quantum
  carry-lookahead adder,'' {\em E-Print: quant-ph/0406142} , 2004.

\bibitem{Shor94}
P.~Shor, ``Polynomial-time algorithms for prime factorization and discrete
  logarithms on a quantum computer,'' {\em 35th Annual Symposium on Foundations
  of Computer Science} , pp.~124--134, 1994.

\bibitem{Aharonov97a}
D.~Aharonov and M.~Ben-Or, ``Fault tolerant computation with constant error,''
  {\em Symposium on Theory of Computing (STOC 1997)} , pp.~176--188.

\bibitem{Nielsen00a}
M.~A. Nielsen and I.~L. Chuang, {\em Quantum Computation and Quantum
  Information}, Cambridge University Press, Cambridge, UK, 2000.

\bibitem{Shor95}
P.~W. Shor, ``Scheme for reducing decoherence in quantum computer memory,''
  {\em Phys. Rev. A}~{\bf 54}, p.~2493, 1995.

\bibitem{Gottesman99}
D.~Gottesman, ``Fault tolerant quantum computation with local gates,'' {\em
  Journal of Modern Optics}~{\bf 47}, pp.~333--345, 2000.

\bibitem{Svore04a}
K.~Svore, B.~Terhal, and D.~DiVincenzo, ``Local fault-tolerant quantum
  computation,'' {\em E-Print: quant-ph/0410047} , 2004.

\bibitem{Yang04}
Y.Yang and J.Wang, ``Pipelined all-to-all broadcast in all-port mesh and
  tori,'' {\em IEEE Transactions on Computers}~{\bf 50}(10), pp.~1020--1032,
  2001.

\end{thebibliography}
